%% file: main.tex
\documentclass[conference, compsoc]{IEEEtran}
\usepackage[utf8]{inputenc}

\usepackage{amsmath,amssymb,amsfonts}
\usepackage{algorithm,algorithmic}
\usepackage{textcomp}
\usepackage{xcolor}
\usepackage{multirow}
\usepackage{soul}
\usepackage{graphicx}
\usepackage{makecell}
\usepackage{microtype}
\usepackage{gensymb}
\usepackage{hyperref}

 \usepackage{enumitem}

\usepackage{tabularx,lipsum,environ}

\makeatletter
\newcommand{\problemtitle}[1]{\gdef\@problemtitle{#1}}
\newcommand{\probleminput}[1]{\gdef\@probleminput{#1}}
\newcommand{\problemquestion}[1]{\gdef\@problemquestion{#1}}
\NewEnviron{problem}{
  \problemtitle{}\probleminput{}\problemquestion{}
  \BODY
  \par\addvspace{.5\baselineskip}
  \noindent
  \begin{tabularx}{\textwidth}{@{\hspace{\parindent}} l X c}
    \multicolumn{2}{@{\hspace{\parindent}}l}{\@problemtitle} \\
    \textbf{Input:} & \@probleminput \\
    \textbf{Question:} & \@problemquestion
  \end{tabularx}
  \par\addvspace{.5\baselineskip}
}
\makeatother

\newcommand{\linebreakand}{%
  \end{@IEEEauthorhalign}
  \hfill\mbox{}\par
  \mbox{}\hfill\begin{@IEEEauthorhalign}
  \and
}

\ifCLASSOPTIONcompsoc
  \usepackage[nocompress]{cite}
\else
  \usepackage{cite}
\fi

\usepackage{graphicx}
\begin{document}

\title{On False Data Injection Attack against Building Automation Systems}

\author{
\IEEEauthorblockN{Michael Cash\IEEEauthorrefmark{1},
Christopher Morales-Gonzalez\IEEEauthorrefmark{2},
Shan Wang\IEEEauthorrefmark{3}\IEEEauthorrefmark{2},
Xipeng Jin\IEEEauthorrefmark{1},
Alex Parlato\IEEEauthorrefmark{1},\\
Jason Zhu\IEEEauthorrefmark{4},
Qun Zhou Sun\IEEEauthorrefmark{1},
Xinwen Fu\IEEEauthorrefmark{1}\IEEEauthorrefmark{2}
}
\IEEEauthorblockA{\IEEEauthorrefmark{1} University of Central Florida. Email: \{mcash001, xipeng\_jin\}@knights.ucf.edu, \{alex.parlato, qz.sun\}@ucf.edu}
\IEEEauthorblockA{\IEEEauthorrefmark{3}Southeast University. Email: shanwangsec@gmail.com}
\IEEEauthorblockA{\IEEEauthorrefmark{4}Email: jjasonzzhu@gmail.com}
\IEEEauthorblockA{\IEEEauthorrefmark{2}University of Massachusetts Lowell. Email: christopher\_moralesgonzalez@student.uml.edu, xinwen\_fu@uml.edu}
}

\maketitle


\begin{abstract}
KNX is one popular communication protocol for a building automation system (BAS). However, its lack of security makes it subject to a variety of attacks. We are the first to study the false data injection attack against a KNX based BAS. We design a man-in-the-middle (MITM) attack to change the data from a temperature sensor and inject false data into the BAS. We model a BAS and analyze the impact of the false data injection attack on the system in terms of energy cost. 
Since the MITM attack may disturb the KNX traffic, we design a machine learning (ML) based detection strategy to detect the false data injection attack using a novel feature based on the Jensen Shannon Divergence (JSD), which measures the similarity of KNX telegram inter-arrival time distributions with {\em attack} and with {\em no attack}. We perform real-world experiments and validate the presented false data injection attack and the ML based detection strategy. We also simulate a BAS, and show that the false data injection attack has a huge impact on the BAS in terms of power consumption. \looseness=-1
\end{abstract}

%
\IEEEpeerreviewmaketitle

\input{sections/sec1_Introduction.tex}
\input{sections/sec2_Background.tex}
\input{sections/sec3_CyberPhysicalAttack.tex}

\input{sections/sec4_Models}
\input{sections/ExtraSection-Defense.tex}
\input{sections/sec6_Evaluation.tex}

\input{sections/sec8_Conclusion.tex}



%

\bibliographystyle{IEEEtran}
\bibliography{IEEEabrv, citation.bib}

\end{document}

%% file: sections/sec1_Introduction.tex
\section{Introduction}
\label{Introduction}

A building automation system (BAS) is one type of cyber-physical system (CPS), which manages and automates a building's mechanical and electrical systems such as heating, ventilation, and air conditioning (HVAC), sensors and actuators. 
Devices in a BAS follow a specific communication protocol, can be linked to a local area network (LAN) and may be reachable from the Internet.
KNX is one popular BAS communication protocol with products market size of \$4.381 billion in 2019 and projected to reach \$10.15 billion by 2026 \cite{360:KNX:2019}.
In this paper, we focus on the KNX based BAS.

As a CPS, the BAS may be subject to both cyber attacks and physical attacks.
A cyber attack may be used to break the networks, pollute, and steal data, while a physical attack may be used to manipulate and damage physical components \cite{wyss_sholander_darby_phelan_2007}. 
Attackers can launch a replay attack against actuators by sniffing traffic messages, so as to manipulate actuators \cite{Antonini2015A_Practical}. 
An adversary may use a hotel room's iPad to access the room's KNX network and identify addresses of KNX devices in hotel rooms. Then attackers can control the KNX devices remotely without the need of the hotel iPad \cite{molina2014}.
Rios presents several password retrieval attacks to exploit and gain access to facility management systems \cite{Rio::FacilityManagementSystems::2014}. 
A fuzzing tool is designed to discover vulnerabilities in the KNX protocol \cite{vacherot:hal-03022310}.
No existing work focuses on the false data injection attack against a KNX based BAS and analyses its impact with regards to energy costs. \looseness = -1

In this paper, we are the first to study the false data injection attack against a KNX based BAS.
Our major contributions can be summarized as follows. We explore physical attacks and cyber attacks to craft a man-in-the-middle (MITM) attack and inject false data to a KNX based BAS. In particular, two Raspberry Pis with KNX adaptors perform as the man in the middle, and relay and change data in KNX telegrams such as temperature sensor data.
We also model a HVAC system and analyze the impact of the fake data injection attack against the BAS in terms of energy cost.\looseness=-1

We design a machine learning based detection strategy to detect the false data injection attack based on statistic features of the traffic, which is disturbed by the MITM attack. The MITM attack cannot be detected by injected false data, which is changed in a minor way.
We design a novel feature which can measure the similarity of the probability distributions of the KNX telegram inter-arrival times with {\em attack} and with {\em no attack} based on Jensen Shannon Divergence (JSD). A ML model is trained with the JSD feature to detect the MITM attack and thus the false data injection attack. \looseness=-1

We perform simulation and real-world experiments to validate the false data injection attack and the ML-based intrusion detection strategy. We simulate the HVAC system under two types of false data injection attacks, and show that the attacks have severe impacts on power consumption.
Our experiments show that the proposed JSD feature works better than common statistics such as mean and variance which are sensitive to outliers inherently. With the SVM-based ML model with the JSD feature, the detection rate reaches 100\% with short KNX telegram segments such as 5 minutes of data. \looseness = -1



\textbf{Responsible disclosure}: We have reported all our findings to Siemens and its development teams. We hope the BAS industry puts more efforts on BAS security.



%% file: sections/sec2_Background.tex
\section{Background}
\label{Background}

In this section, we introduce the communication mediums, address types and telegrams in the KNX protocol.

\subsection{Communication Medium}

The KNX protocol supports multiple physical communication mediums such as twisted pair (TP1), KNXnet/IP, radio frequency (RF), and powerline (PL) for connected devices to exchange information via a KNX network. 
The KNX network can be configured in either a tree, star, or line (daisy-chained) topology. 
With the TP1 backbone communication medium, the KNX network is a bus network. Any message that is sent onto the bus network will be received by every device on the network. 

\vspace{-1mm}
\subsection{Address Type}

The KNX protocol supports two general types of 16-bit addresses: individual addresses and group addresses, which can be used to access devices and data in a KNX network.

\textbf{Individual Address}.
An individual address uniquely identifies a KNX device on a KNX network. 
It has three parts in the notation \texttt{"area.line.deviceAddress"}.
The individual address whose \texttt{deviceAddress} is 0 is reserved for a line coupler. 
If \texttt{deviceAddress} is not 0, the address represents a regular KNX device.
Individual addresses allow devices to access other devices on the network.
For example, when a device sends a {\em read} request, the source address in the request data frame contains the individual address of the device. 
In the response of the request, the destination address will be the individual address of the device which initially sent the request.


{\bf Group Address}:
A group address is a logical entity that is linked to a specific piece of data, called a KNX object, which a KNX device has to provide.
A group address can be either 2 levels \texttt{"main/sub"}, or 3 levels \texttt{"main/middle/sub"}.
Please note that the addresses ``0/0" and ``0/0/0" are reserved for the broadcast addresses.
The linking of a KNX object to a group address allows the object to be read from or written to through the address.
For example, when a device reads a KNX object, the destination address of the {\em read} request will be the group address which is linked to the KNX object.

\vspace{-1mm}
\subsection{Telegram}
\label{sec::telegram}

\begin{figure*}
    \centering
    \includegraphics[width=\textwidth]{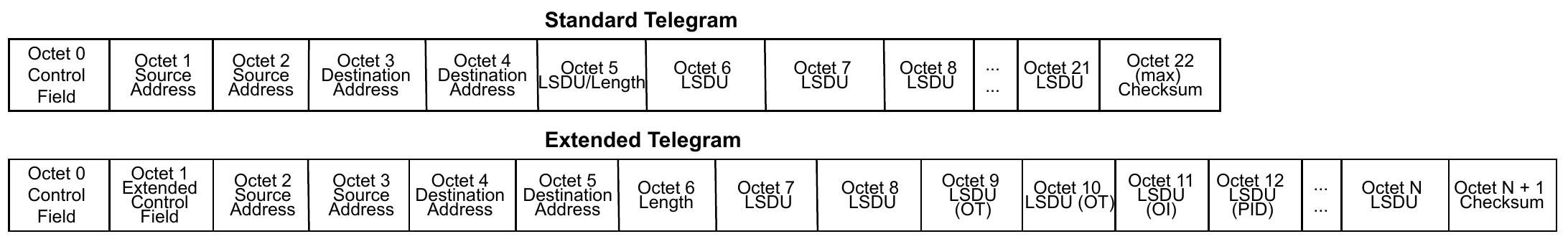}
    \caption{Standard vs Extended Telegrams}
    \label{fig:LTEvsStd}
    \vspace{-2mm}
\end{figure*}

The data frames that are sent throughout a KNX network are known as {\em telegrams}.
The structure of these telegrams will change dependent on both the communication medium and the programming mode in which the KNX network is configured in.
A KNX network with the TP1 communication medium supports several types of programming modes such as standard mode (S-mode) and easy mode (E-mode).
S-mode is the most common mode, in which devices and objects are manually added and configured. 
A variation of E-mode called logical tag extended mode (LTE-mode) is specifically developed for HVAC applications.
In LTE-mode, the configuration of a device is automatically performed by a pre-programmed controller.
Figure \ref{fig:LTEvsStd} shows the structures of standard telegrams in S-mode and extended telegrams in LTE-mode. A specific bit in the {\em control field} of a telegram indicates the telegram type. 


{\bf Standard Telegrams in S-mode}.
Standard telegrams in S-mode are the most common telegrams. 
The {\em control field} in a standard telegram as shown in Figure \ref{fig:LTEvsStd} contains a plethora of information such as the repetition status, priority level and telegram type, which will be used in the interpretation of the telegram.
The {\em source address} details the origin of the telegram while the {\em destination address} specifies the desired recipient. 
The data link layer service data unit ({\em LSDU}) contains the information about the purpose of the telegram such as \textit{groupRead} or \textit{groupWrite}, which will read from or write to a group address respectively. In the case of a \textit{groupWrite}, the LSDU will also contain the data that is to be written. 
The last field {\em checksum} is used to verify the integrity of the telegram.


{\bf Extended Telegrams in LTE-mode}.
Extended telegrams in LTE-mode are larger than standard telegrams due to the inclusion of the {\em extended control field} and a larger LSDU as shown in Figure \ref{fig:LTEvsStd}. 

The extended control field in an extended telegram can be divided into three parts: the address type, hop count, and extended frame format (EFF). 
The address type indicates the type of the destination address, which may be an individual address or a group address.
The hop count limits the telegram distance to ensure that the telegram does not infinitely traverse the network. 
If the address type is a group address, then the EFF will describe how to interpret the group address. Otherwise, it will interpret the destination address as a normal individual address.

A LTE device is presumed to have multiple objects that can be of the same type. As a result, a larger LSDU is needed to access a specific object of the LTE device. 
When reading a KNX object of a LTE device, 
the object type (OT), object index (OI), and a property ID (PID) all must be provided.
These are additional fields located in the {\em LSDU} field of an extended telegram. These three fields indicate a particular object and describe a property that a device desires to read. 
A tag address is also needed to read a KNX object of a LTE device. The tag address can be interpreted from a group address based on the EFF described previously. 

%% file: sections/sec3_CyberPhysicalAttack.tex
\vspace{-1mm}
\section{False Data Injection Attack}
\label{CyberPhysicalAttacks}

In this section, we first introduce physical attacks against KNX that will allow an attacker to gain access into a KNX network.
We then present an eavesdropping attack, which can be used to learn detailed protocols of the KNX based BAS. 
Next, we present the false data injection attack via a man-in-the-middle (MITM) attack. 
Finally, we model a KNX based BAS, i.e. a HVAC system, and formally analyze the impact of false data injection attacks on the HVAC system in terms of energy cost.

\vspace{-1mm}
\subsection{Physical Attack}
A KNX device such as a temperature sensor is often placed in publically accessible locations and thus is subject to a physical attack. 
For a KNX device utilizing the communication medium TP1, attackers can remove the device from the KNX network, i.e. a bus network, by removing its twisted pair cables from either the bus end or the device end. 
If the cables are disconnected from the device end, the same cables can be used by a malicious device to gain access to the KNX network. 
A KNX device may have four terminal ports and each port can be used to connect to another KNX device.
If one of the four ports is available, a complete disconnect of the TP1 cables is not required. A malicious device can utilize the available port to connect and access to the KNX network. Then the malicious device can see any telegrams being transmitted in the network, or send its own telegrams to the network. 

\vspace{-1mm}
\subsection{Eavesdropping Attack}

In the eavesdropping attack, the attacker attaches their own device to the KNX network through the described physical attack above, and then passively listens on the bus for any telegrams that are sent.
It provides a means of understanding how the KNX telegrams of a device are formed and the detailed communication protocol between devices.
After performing a data dump, the attacker is presented with a plethora of information such as: source and destination addresses and individual/group addresses. In the case of extended telegrams, tag addresses, OTs, OIs and PIDs are also presented. With this gained information, the attacker can craft their own telegrams with respect to the victim device. 

\vspace{-1mm}
\subsection{False Data Injection via MITM Attack}

\begin{figure}
    \centering
    \includegraphics[width=1\linewidth]{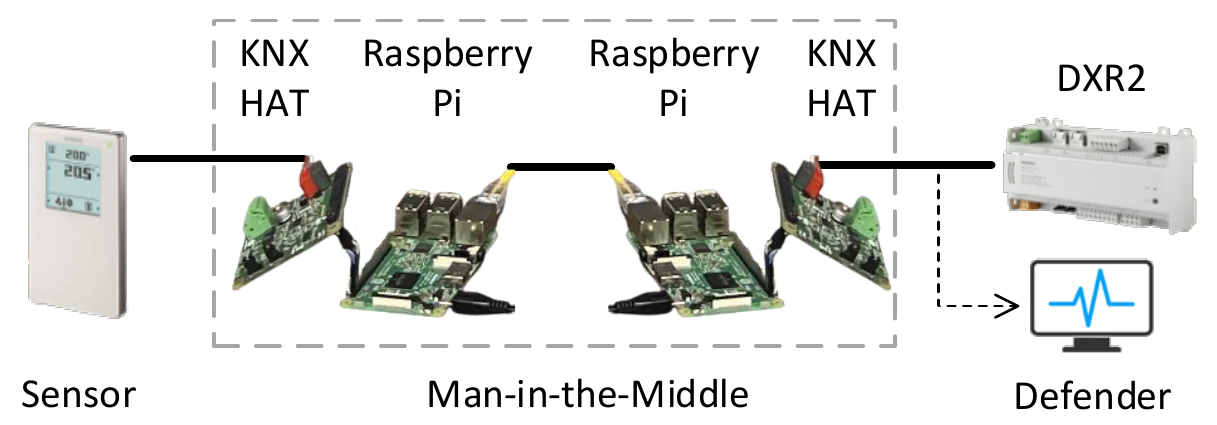}
    \caption{MITM Attack against Sensors
    }
    \label{fig:MITM}
    \vspace{-2mm}
\end{figure}

With the information gained from eavesdropping and properly crafted telegrams, 
attackers can deploy MITM attacks as shown in Figure \ref{fig:MITM} against a KNX based BAS with the goal of false data injection. 
Without loss of generality,
we use the temperature sensor as a victim device in the MITM attack example.
Two Raspberry Pis are used to perform the MITM attack.
The left Pi is connected to the KNX network that the temperature sensor is connected to via a KNX adaptor, KNX Pi Hat. It then changes the data within the {\em LSDU} of received telegrams from the sensor, and forwards these changed telegrams to the right Pi via Ethernet or wirelessly. The right Pi is connected to the KNX network that the DXR2 controller is on, and sends the changed telegrams to the DXR2. The two Pis also forward telegrams generated by the DXR2 to the victim temperature sensor and other types of telegrams, excluding the temperature readings generated by the victim sensor, to the DXR2 without modification.

Using two Raspberry Pis is necessary for stealthy data injection. One Pi can use only one KNX adaptor in out setup.
If one Pi and one KNX adaptor are used in the MITM attack, the victim temperature sensor will be connected to the DXR2 directly through the KNX adaptor. Although the Pi can still gather the temperature sensor's telegrams, change them, and forward them to the DXR2, the DXR2 will receive both the original and changed telegrams; This would lead a defender monitoring the DXR2's KNX network to easily detect the attack.
Using two Pis and two KNX adaptors separates the temperature sensor and the DRX2 into two different KNX networks and addresses the issue above.


%% file: sections/sec4_Models.tex
\subsection{Impact of False Data Injection}
\label{sec::model}
We can utilize the MITM attack to control various KNX devices such as a temperature sensor, and to inject fake data into a HVAC system. In this paper, we focus on injecting false room temperature values by manipulating a temperature sensor, which may incur severe energy costs.
Next, we model a HVAC system. Then, we formalize the power consumption incurred by room temperature changes in terms of cooling in hot weather, so as to quantify the energy cost incurred by injecting false room temperature values. \looseness = -1

\begin{figure}[ht]
    \centering
    \includegraphics[width=1.0\linewidth]{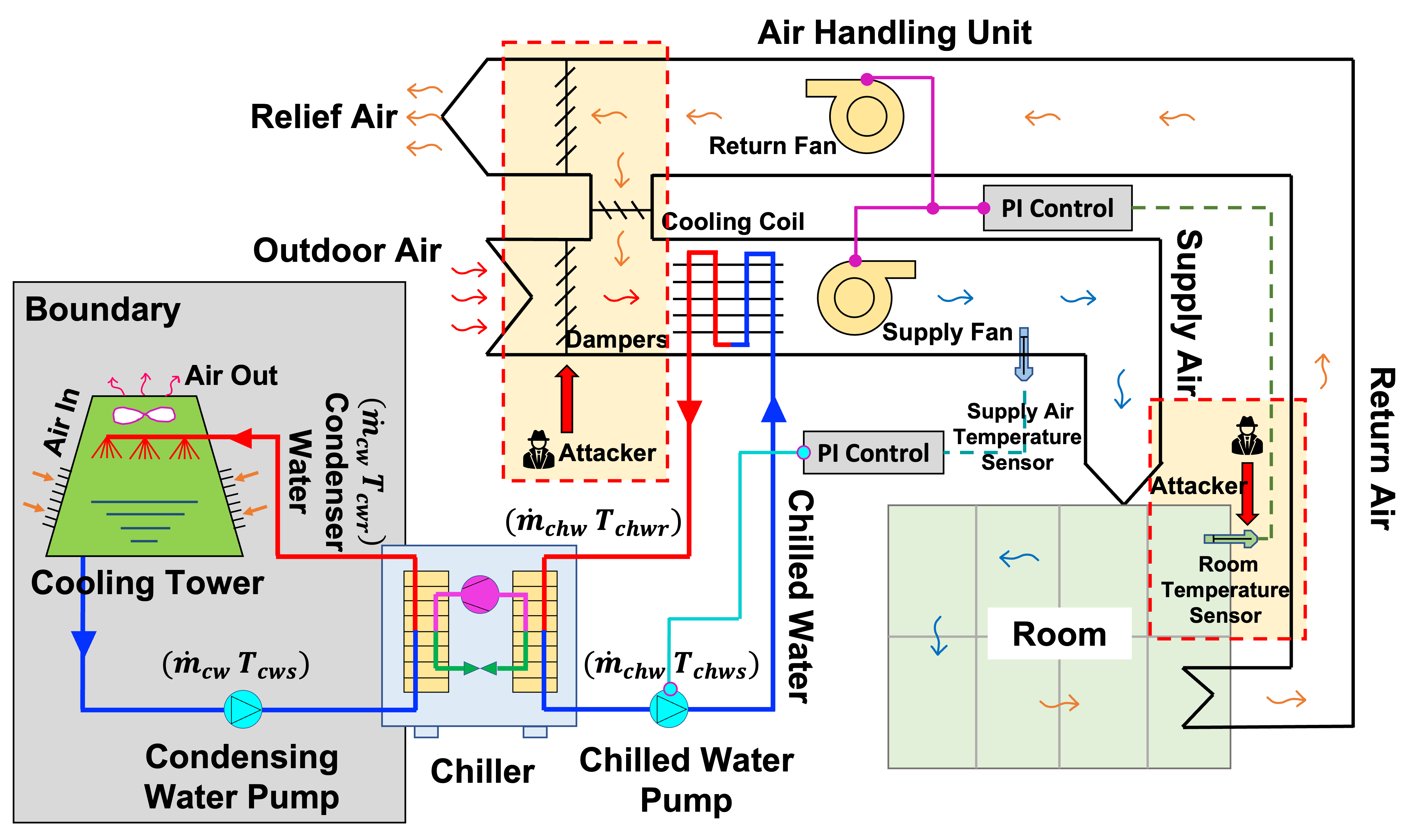}
    \vspace{-2mm}
    \caption{HVAC system model}
    \label{fig:HVAC_structure}
    \vspace{-2mm}
\end{figure}

We model a HVAC system as shown in Figure \ref{fig:HVAC_structure}, which consists of a chiller that removes heat from water---the coolant, the chilled water pump, air handling unit (AHU) with three dampers and two fans which circulate the air, and the constant boundary conditions which represent a typical cooling tower system. The AHU serves one thermal zone such as a room.
These components work together to maintain the room temperature at a predefined setpoint.
Assume the room temperature increases. Then components in the HVAC system will work together to cool down the room temperature to maintain it at the setpoint, which incurs power consumption of fans, the pump and the chiller.


The total power consumption of the HVAC system is the summation of the power consumed by fans, the chilled water pump, and the chiller as follows \cite{stewart2018surface,wang2022control,hydeman2002tools}.
\begin{equation}
\label{eqn::ptotal}
\begin{split}
P_{total}(t) &=  P_{fan}(T_r(t)) \\
 & + P_{pump}(T_{sa}(t)) \\ 
 & + P_{chiller}(\dot{m}_{chw}(t)).
\end{split}
\end{equation}
$T_r(t)$ is the room temperature and may change with time. The power consumption of fans $P_{fan}(.)$ increases with the room temperature. $T_{sa}(t)$ is the supply air temperature and increases with $T_r(t)$. The power consumption of the pump $P_{pump}(.)$ increases with $T_{sa}(t)$ and thus the room temperature. $\dot{m}_{chw}(t)$ is the chilled water mass flow rate and increases with $T_{sa}(t)$. So the power consumption of the chiller $P_{chiller}(\dot{m}_{chw}(t))$ increases with the room temperature. \looseness=-1

It can be observed from Equation (\ref{eqn::ptotal}) and discussion above that when the room temperature increases, the power consumption by fans, the chilled water pump, and the chiller increases and thus the total cooling power consumption increases. Therefore, when an attacker uses the false data injection attack to change the room temperature reading, the cooling system may be activated and the attack may incur energy waste and cost. 

%% file: sections/ExtraSection-Defense.tex
\section{Defense}

In this section, we first present the detection problem definition, and show the traffic patterns with {\em attack} and with {\em no attack} are different. We then discuss selection of features and classifier for machine learning. At last, we introduce the attack detection with trained machine learning models.


\subsection{Problem Definition for False Data Injection Attack Detection}

The false data injection attack presented in Section \ref{CyberPhysicalAttacks} utilizes the MITM attack. We carefully code the MITM attack to reduce potential delays in forwarding KNX telegrams by Pis. However, the attack may still change the KNX network traffic pattern due to network delays and jitters. In order to detect the false data injection, we can detect the MITM attack. For the detection, we train a machine learning (ML) model utilizing traffic pattern features. We assume that a defender collects telegram data and performs detection at the DXR2 controller as shown in Figure \ref{fig:MITM}.

The raw data we collect is telegrams with timestamps, from which we derive the telegram arrival time data series. From two consecutive telegrams arriving at $t_i$ and $t_{i+1}$, we can derive the telegram inter-arrival time $I_i=t_{i+1} - t_i$. Therefore, from a telegram arrival time data series, we can derive a telegram inter-arrival data series.

We use a machine learning model to detect the false data injection attack. For training purposes, the defender collects two labeled telegram inter-arrival time training data sets : $A$=$\{ I_{A, i} ~|~ 0 < i \le l \}$, which is the telegram inter-arrival time data set when there is an attack; $B$=$\{ I_{B, i} ~|~ 0 \le i < m \}$, which is the telegram inter-arrival time data set when there is no attack. 
We define the {\em detection time window} $t$ as the time duration of the telegram inter-arrival time data segment, which the defender collects for attack detection. \looseness = -1

{\bf Problem Definition}: 
Given training data $A$ and $B$ and a collected test telegram inter-arrival time segment $\mathcal{T}$ of a time duration of $t$, does $\mathcal{T}$ belong to {\em attack} or {\em no attack}?

\begin{figure}[ht]
\centering
\begin{minipage}{0.45\columnwidth}
  \centering
\includegraphics[height=0.66\textwidth]{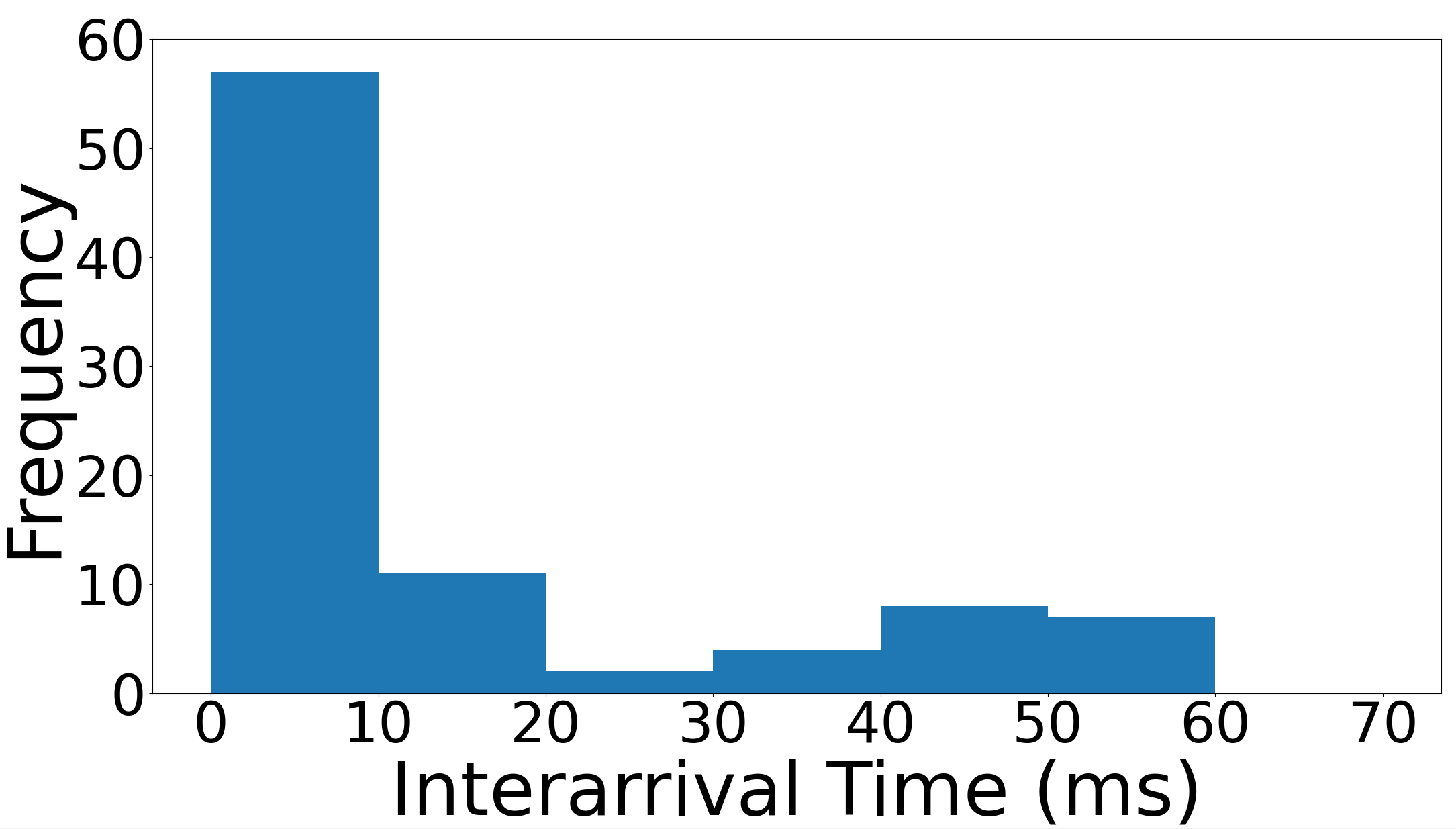}
    \vspace{-2mm}
  \caption{Histogram of one segment in {\em attack} dataset $A$ with 20min detection time window}
  \label{fig:subsetattack1}
    \vspace{-2mm}
\end{minipage}
~~~
\begin{minipage}{0.45\columnwidth}
  \centering
\includegraphics[height=0.66\textwidth]{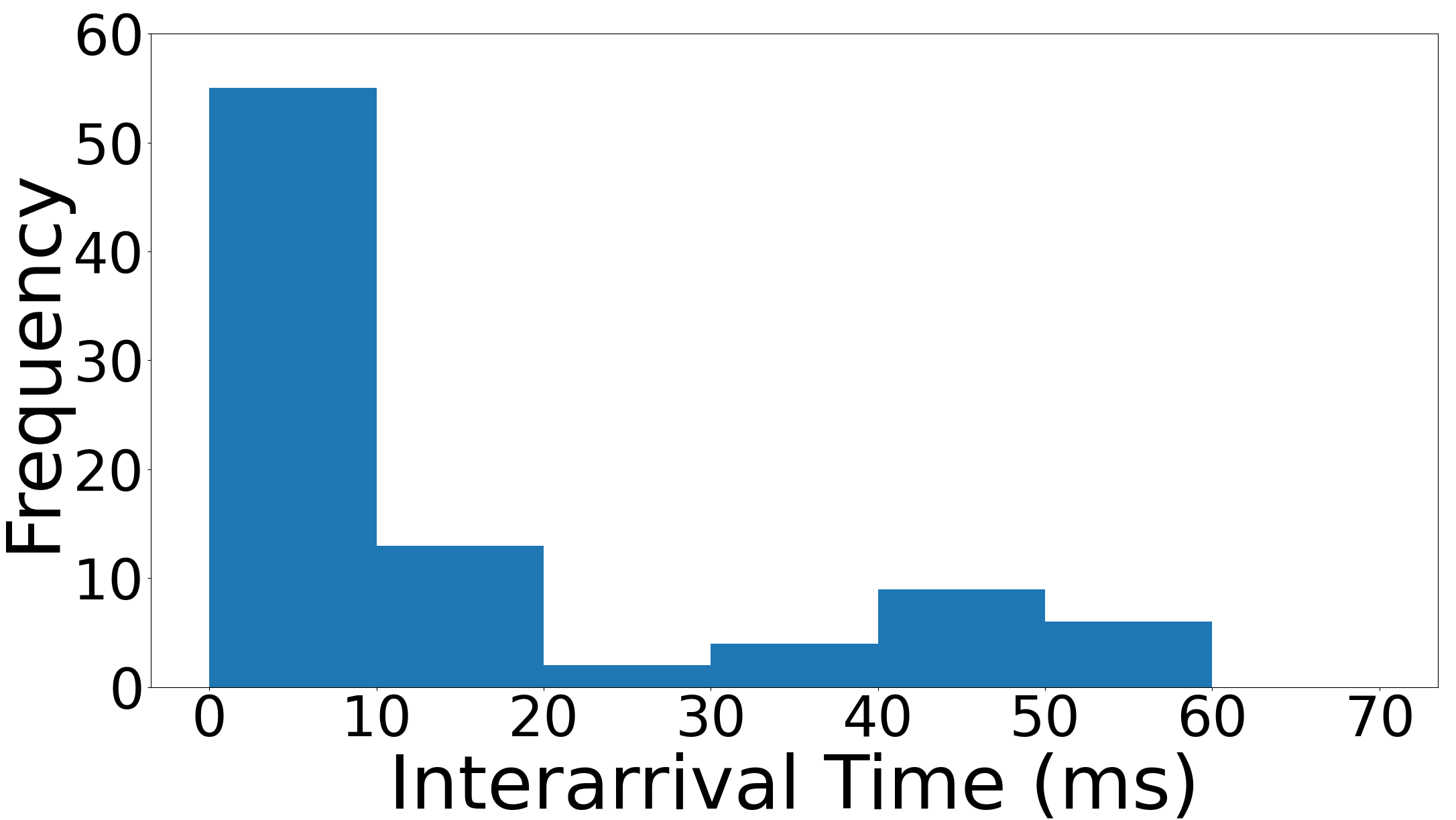}
    \vspace{-2mm}
  \caption{Histogram of another segment in {\em attack} dataset $A$ with 20min detection time window}
  \label{fig:subsetattack2}
    \vspace{-2mm}
\end{minipage}
\end{figure}
\begin{figure}[ht]
\centering
\begin{minipage}{0.45\columnwidth} 
  \centering
\includegraphics[height=0.66\textwidth]{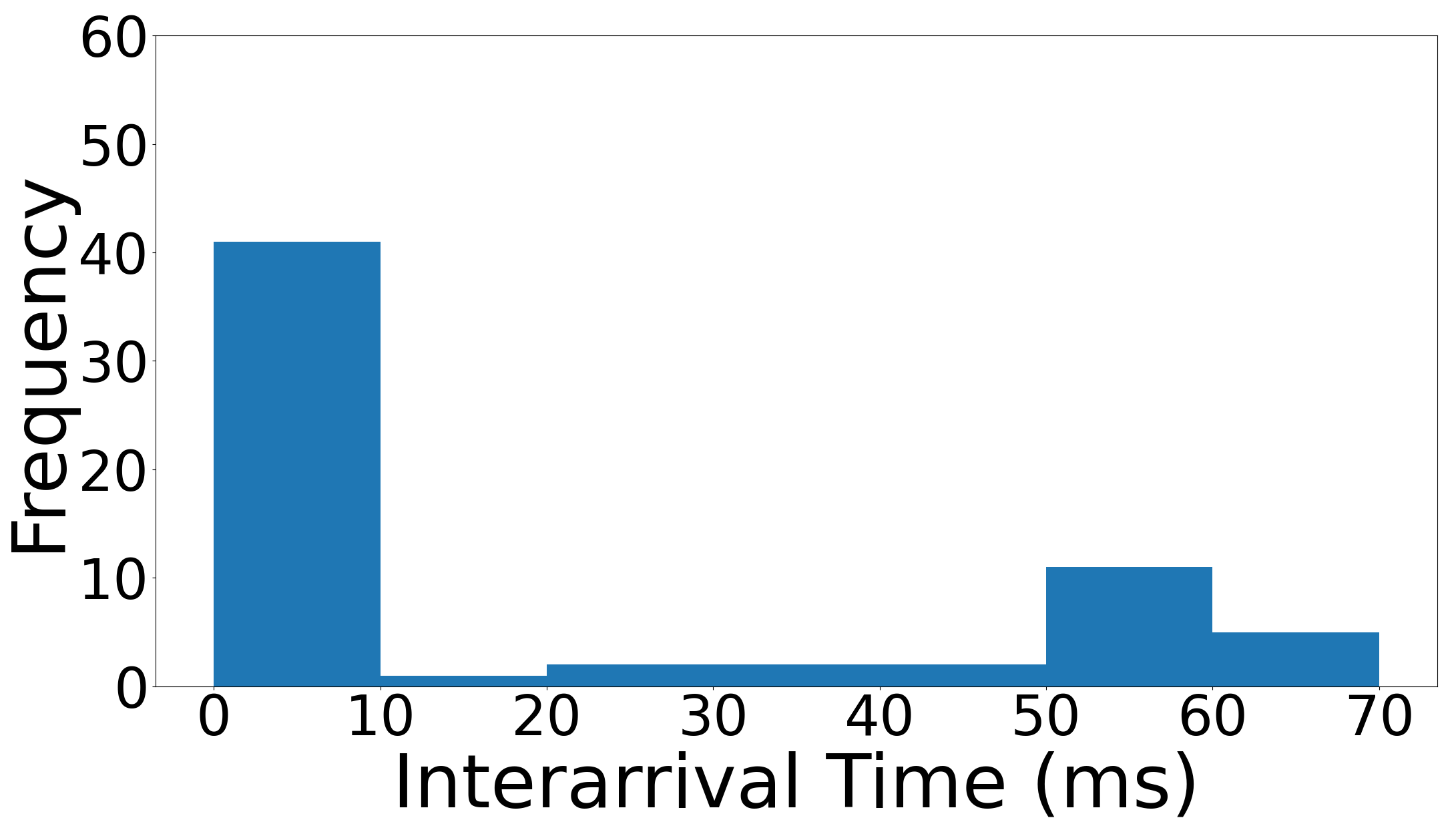}
    \vspace{-2mm}
  \caption{Histogram of one segment in {\em no attack} dataset $B$ with 20min detection time window}
  \label{fig:subsetbaseline1}
    \vspace{-2mm}
\end{minipage}%
~~~~
\begin{minipage}{0.45\columnwidth}
  \centering
  \includegraphics[height=0.66\textwidth]{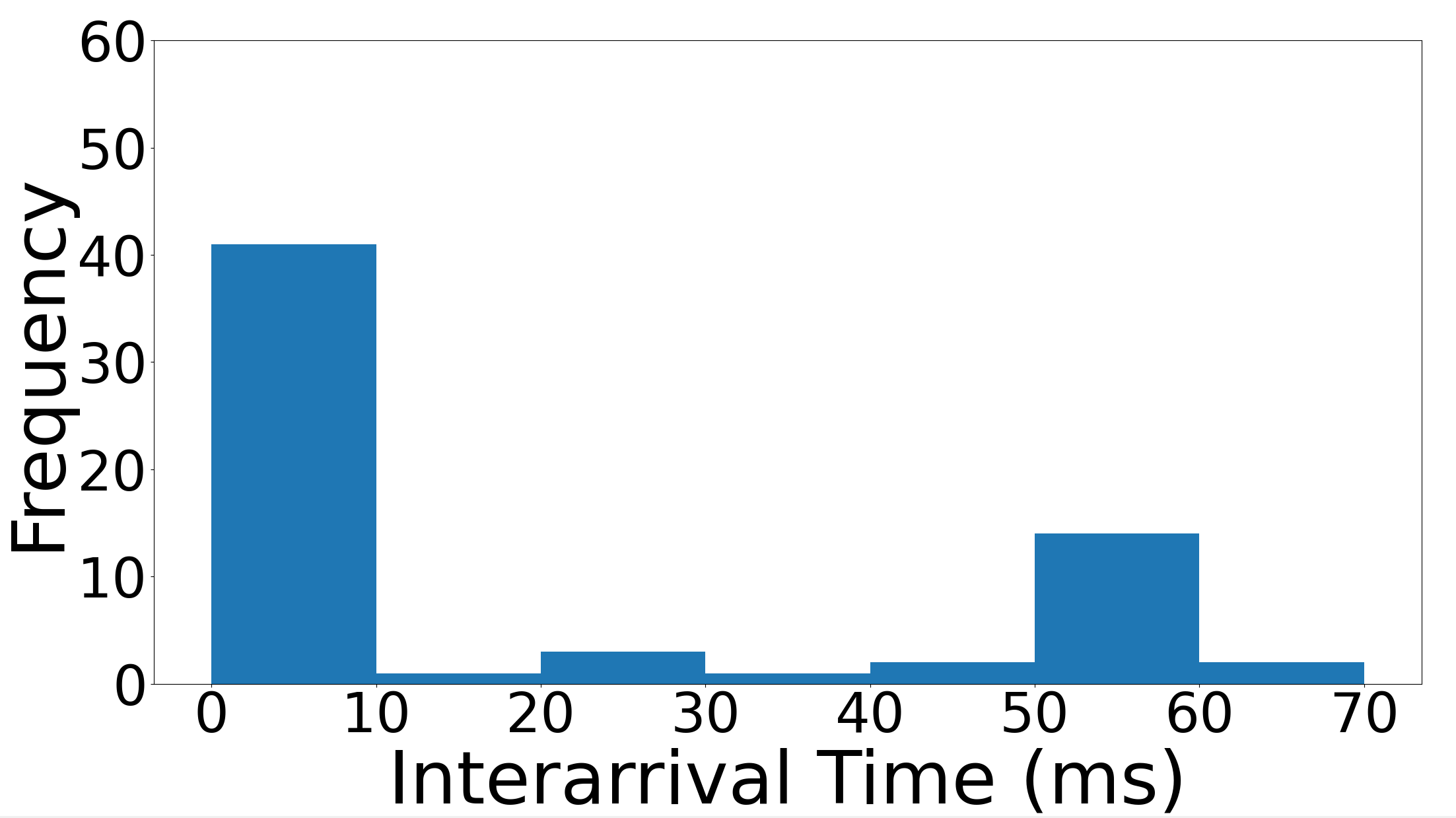}
    \vspace{-2mm}
  \caption{Histogram of another segment in {\em no attack} dataset $B$ with 20min detection time window}
  \label{fig:subsetbaseline2}
    \vspace{-2mm}
\end{minipage}
\end{figure}

\subsection{Intuition of Attack Detection}

We want to observe the patterns of telegram inter-arrival times with attack and with no attack, and see if there is any difference. 
We draw the histograms of telegram inter-arrival time segments of a detection time window in training data sets $A$ and $B$, and observe the distribution of the inter-arrival time. 
For a detection time window of 20 minutes, Figures \ref{fig:subsetattack1} and \ref{fig:subsetattack2} show the histograms of two segments in $A$ with attack.
Figures \ref{fig:subsetbaseline1} and \ref{fig:subsetbaseline2} show the histograms of two segments in $B$ with no attack. 
It can be observed that the histograms from the same data set are similar while the histograms from $A$ and $B$ look different. 
For example, there are more small telegram inter-arrival times in $A$ than in $B$.
Therefore, the telegram inter-arrival time with attack has different patterns from the telegram inter-arrival time with no attack. 

\vspace{-1mm}
\subsection{Feature Selection}
We divide the telegram inter-arrival time training data sets $A$ and $B$ into segments of a time duration of $t$. For $A$, assume we obtain $u$ segments $\{ S_{A,i} ~|~ 0 \le i < u \}$. For $B$, we derive $v$ segments, $\{ S_{B, j} ~|~ 0 \le j < v \} $.

\subsubsection{Mean and Variance as Features}
Since Figures \ref{fig:subsetattack1}-\ref{fig:subsetbaseline2} show telegram inter-arrival times with attack and with no attack have different patterns, the first features we consider are the $mean$, $variance$ of the telegram inter-arrival time segment and the vector $(mean, variance)$. 
For our telegram inter-arrival time data, we derive the feature $mean$ as follows. For each telegram inter-arrival time segment in $A$ and $B$, we can derive its mean. Therefore, we derive the training data sets of the feature $mean$ as follows, $\{ \mathcal{M}_{A,i} ~|~ 0 \le i < u \}$
and $\{ \mathcal{M}_{B, j} ~|~ 0 \le j < v \}$. We also derive the mean for the test data $\mathcal{T}$ as $\mathcal{M}_\mathcal{T}$. 
Similarly, we can derive the training data sets of the feature variance as $\{ \mathcal{V}_{A,i} ~|~ 0 \le i < u \}$
and $\{ \mathcal{V}_{B, j} ~|~ 0 \le j < v \}$ and the variance for the test data $\mathcal{T}$ as $\mathcal{V}_\mathcal{T}$. For the feature $(mean, variance)$, we then have $\{ (\mathcal{M}_{A,i}, \mathcal{V}_{A,i}) ~|~ 0 \le i < u \}$, $\{ (\mathcal{M}_{B, j}, \mathcal{V}_{B, j}) ~|~ 0 \le j < v \}$ and $(\mathcal{M}_\mathcal{T}, \mathcal{V}_\mathcal{T})$.


\subsubsection{Probability Distribution Similarity as Feature}
Since mean and variance are inherently sensitive to outliers and not robust as features, we want to find a feature based on the probability distribution given that the outliers have a small probability to occur and features considering probabilities are more robust.
We propose to use the Jensen Shannon Divergence (JSD) to measure the similarity between probability distributions of telegram inter-arrival time segments. 

We derive the JSD similarity of two telegram inter-arrival time segments as follows. From the two segments, we can derive their corresponding telegram inter-arrival time distributions $P$ and $Q$. $\mathcal{X}$ is the range of the telegram inter-arrival time.
\begin{equation}
KL(P||Q) = \sum\limits_{x \in \mathcal{X}}P(x)log\frac{P(x)}{Q(x)}
\end{equation}
\vspace{-2mm}
\begin{equation}
JSD(P||Q) = \frac{1}{2}KL(P||M) + \frac{1}{2}KL(Q||M),
\end{equation}
where $M = \frac{1}{2}(P+Q)$,
and $JSD \in [0, 1]$. The closer to 0 $JSD$ is, the more similar these two segments are. 

We can now derive JSD features for our training data sets $A$ and $B$, and the test data $\mathcal{T}$.
\begin{itemize}
\item For each segment $S_{A,i}$ ( $0 \leq i < u$ ) of $A$, we measure its JSD similarity with all segments of $B$ and derive a feature vector of $v$ elements as follows
\begin{equation}
\left( JSD(P_{A,i} || P_{B,0}), \cdots,  JSD(P_{A,i} || P_{B,v-1} \right).
\end{equation}
Therefore, we have $u$ feature vectors as training data. Those feature vectors are the {\em attack} feature vectors, which measure the similarity between {\em attack} data and {\em no attack} data.

\item We now derive the {\em no attack} feature vectors for training. For each segment $S_{B,j}$ ( $0 \leq j < v$ ) of $B$, we measure its JSD similarity with all segments of $B$ and derive a feature vector of $v$ elements as follows
\begin{equation}
\left( JSD(P_{B,j} || P_{B,0}), \cdots,  JSD(P_{B,j} || P_{B,v-1} \right).
\end{equation}
Therefore, we have $v$ {\em no attack} feature vectors.

\item We derive the feature vector for test data $\mathcal{T}$ as follows
\begin{equation}
\left( JSD(P_\mathcal{T} || P_{B,0}), \cdots,  JSD(P_\mathcal{T} || P_{B,v-1} \right).
\end{equation}
\end{itemize}

\vspace{-1mm}
\subsection{Classifier}
For each candidate feature including $mean$, $variance$, $(mean, variance)$ and JSD similarity,
we have derived two training data sets of feature vectors and the test data feature vector. Note the features $mean$ and $variance$ can be viewed as feature vectors with one element.
Therefore, for each type of feature vector,
we can train a classification model to classify the test data as {\em attack} or {\em no attack}. In our experiments, we use 70\% of the data we collect for the training purpose and the other 30\% as test data. 
We label both the training data and the test data with a class based on the ground truth.


Two classic ML algorithms, i.e. decision tree and support vector machine (SVM), are used to train binary classification models. 
The decision tree classifier is a tree-like model of decisions. Each internal node is a test on a feature in a feature vector, and each branch of this internal node is an output of the test. A leaf node is a {\em no attack} class or an {\em attack} class. Given an input feature vector, the decision tree model finds a path from the root to a leaf node based on the value of each feature, and uses the class of the leaf node as the output.  
The SVM based classifier is a function which splits a space into parts with different classes. Given an input feature vector, the function maps the feature vector to a subspace and outputs the class of the subspace.

\vspace{-1mm}
\subsection{Attack Detection}

We use a trained ML model to detect the MITM attack, thus the false data injection attack. 
For real-time detection, we can just collect a telegram arrival time series in a period of $t$ that the ML model uses, calculate its inter-arrival time segment $\mathcal{T}$, derive the feature vector for $\mathcal{T}$ per the features that the ML model uses, and then input the feature vector to the trained ML model which will output a class for $\mathcal{T}$.
We use the feature vectors of segments in the testing data set to evaluate the detection rate. 
By comparing the output class with the true class label of a testing feature vector, we can know if the classification, i.e. attack detection, is correct, and obtain the detection rate.
In our experiments, we set different detection time windows $t$ and evaluate the detection rate of different ML models with different features versus $t$. \looseness = -1


%% file: sections/sec6_Evaluation.tex
\section{Evaluation}
\label{Evaluation}

In this section, we first present the experiment setup. Then we 
show the impact of the false data injection attack on the HVAC system.
Finally, we demonstrate the effectiveness of the proposed ML-based defense scheme.

\subsection{Experiment Setup}

\begin{figure}[ht]
    \centering
    \includegraphics[width=1\linewidth]{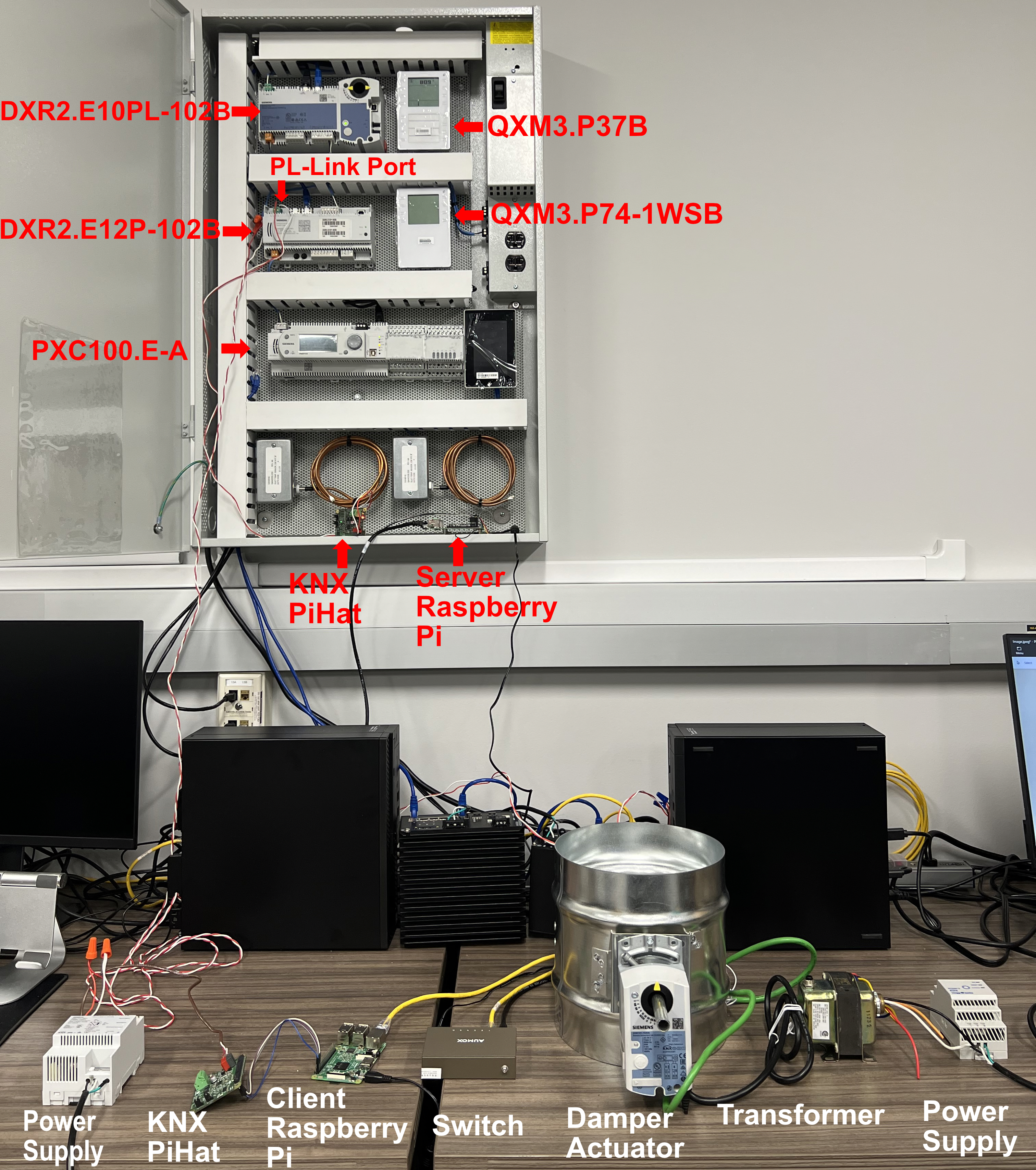}
    \caption{Experiment setup for MITM attack against a temperature sensor}
    \label{fig:fieldPanel}
    \vspace{-2mm}
\end{figure}

Figure \ref{fig:fieldPanel} shows the experiment setup using a Siemens BACnet Field Panel located at the University of Central Florida. We use the Siemens DXR2.E12P-102B (DXR2) BAS controller and the Siemens QMX3.P74B-1WSB Room Operator Unit, which can work as a KNX temperature sensor in our experiments. The KNX device data from the DXR2 can be viewed and monitored in real time by the Siemens Desigo CC building management software platform. We use two Raspberry Pi 3s with 16 GB of flash memory running Raspbian OS version 10-Buster. 
We use two KNX Raspberry Pi HATs (PiHAT), each of which is connected to a Pi's universal asynchronous receiver/transmitter (UART) on-board pins using 4 Dupont wires. \looseness=-1

We use twisted pair (TP1) cables to connect one Pi HAT to the DRX2, forming one KNX network, and connect the second Pi HAT to the temperature sensor to form another KNX network. 
An Ethernet switch is used to connect the two Pis together so that the two Pis may communicate with each other to perform the MITM attack. 
We use the Calimero Java library for KNX  
to construct telegrams, perform eavesdropping and forward telegrams. \looseness=-1

We validate the MITM attack against the temperature sensor as shown in Figure \ref{fig:fieldPanel}. Please note that the MITM attack against the damper is also performed but not presented in this paper.

\begin{figure}[ht]
  \centering
\includegraphics[width=0.75\linewidth]{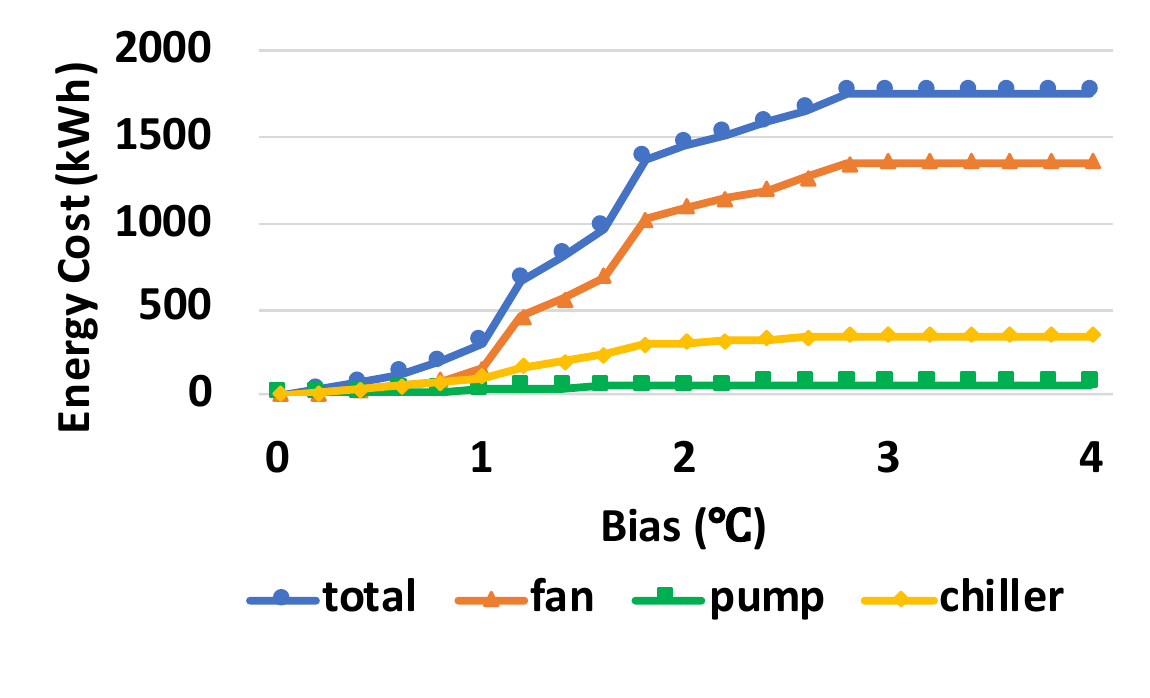}
    \vspace{-2mm}
  \caption{Additional energy cost incurred by Attack-i}
  \label{fig::TRoomBias}
    \vspace{-2mm}
\end{figure}
\begin{figure}[ht]
  \centering
  \includegraphics[width=0.75\linewidth]{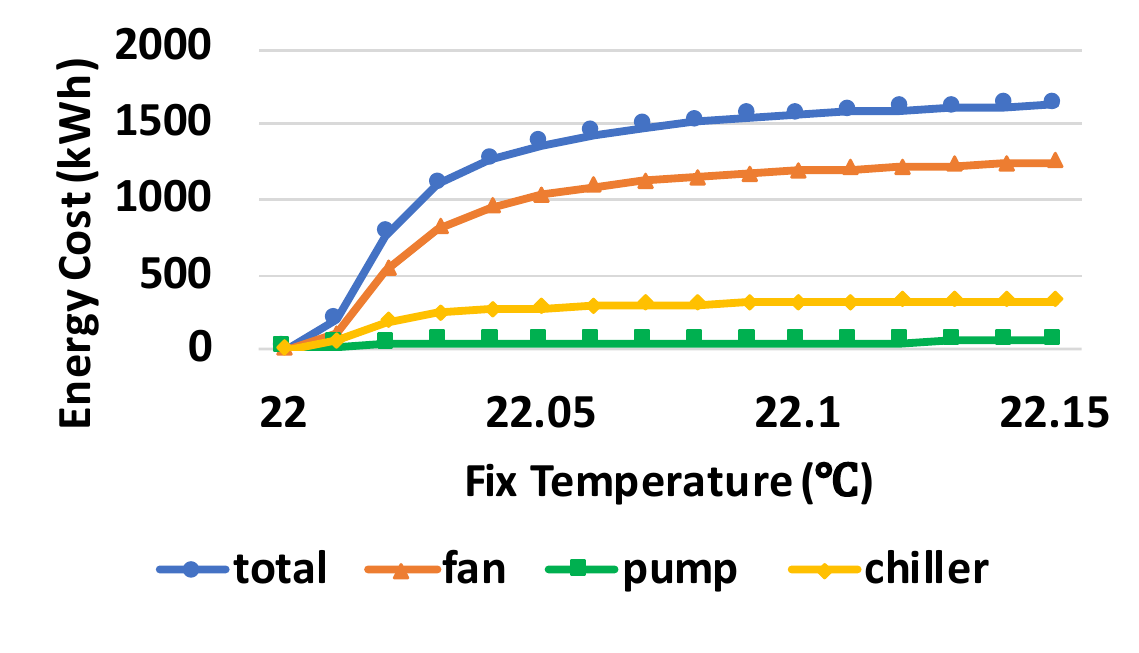}
    \vspace{-2mm}
  \caption{Additional energy cost incurred by Attack-ii}
  \label{fig::TRoomFix}
    \vspace{-2mm}
\end{figure}

\subsection{Attack Impact on HVAC System}
\label{sec::simulation}

We use a simulation tool named $Dymola$ and the Modelica buildings library \cite{wetter2014modelica} to simulate the HVAC system model as introduced in Section \ref{sec::model} and mimic the real-time attacks to evaluate the impact of the false data injection attack. \looseness = -1


{\bf Simulation Settings.}
In the simulation, a built-in solar radiation module of $Dymola$ is used to simulate solar radiation (i.e., sun) influence to the building. 
We use the TMY3 weather data of Orlando, FL on July 29, 2005 to obtain a realistic ambient (i.e., surrounding) air temperature, humidity and wind speed.
The room temperature setpoint, the supply air temperature setpoint and the supply water temperature setpoint are 22$^{\circ}$C, 14$^{\circ}$C and 6$^{\circ}$C respectively. 
The damper opening level is set as 30\%.
The room size is 29,750 cubic meters.
In the simulation, the outside temperature changes along with time, which affects the room temperature and incurs energy cost. Without attack, the energy cost incurred by the outside temperature changes is 1,303.70 kWh from 7:00 am to 7:00 pm. We use the 1,303.70 kWh as the energy cost benchmark.
We evaluate two types of false data injection attacks as follows. 
(Attack-i) Constantly adding a fixed bias such as 1$^{\circ}$C to the original room temperature sensor reading.
(Attack-ii) Overwriting the original room temperature sensor reading to a constant false value such as 22.005$^{\circ}$C which is higher than the setpoint.
We launch each attack at 7:00 am and cease the attack after 12 hours.

{\bf Energy Cost.}
Figures \ref{fig::TRoomBias} and \ref{fig::TRoomFix} show the additional energy cost incurred by the two false data injection attacks in comparison to the energy cost benchmark 1,303.70 kWh without attack. For each attack strategy, we measure the total additional energy cost and the additional energy cost of the fans, the chilled water pump and the chiller.
It can be seen that the power consumption of the fans, pump and chiller has risen significantly. 
Our proposed attack strategies can have a huge impact on HVAC system power consumption.

\begin{figure}[ht]
    \centering
    \includegraphics[width=0.8\linewidth]{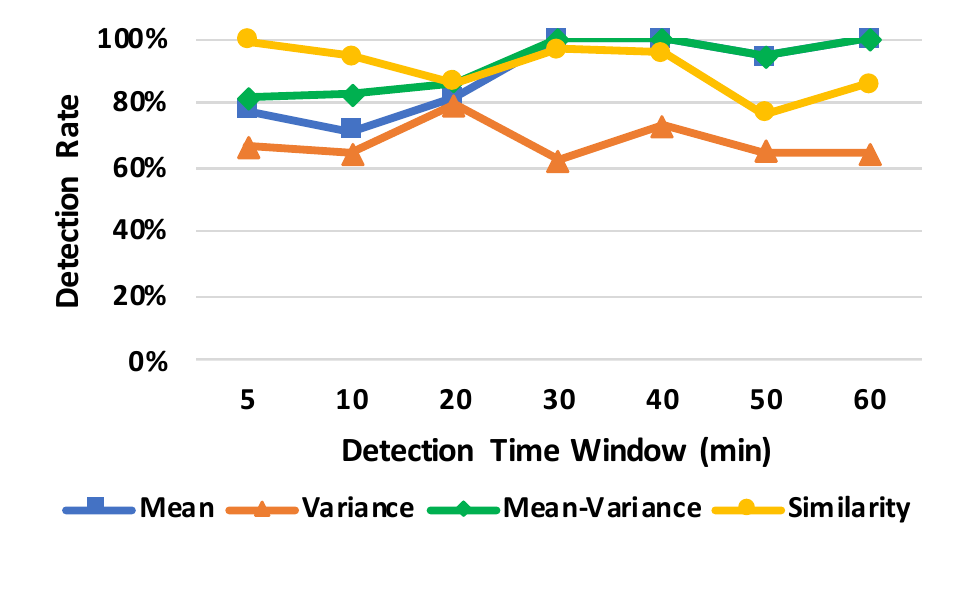}
    \vspace{-2mm}
    \caption{Detection Rates vs. Detection Time Window for J48}
    \label{fig:detectionratej48}
    \vspace{-2mm}
\end{figure}

\begin{figure}[ht]
    \centering
    \includegraphics[width=0.8\linewidth]{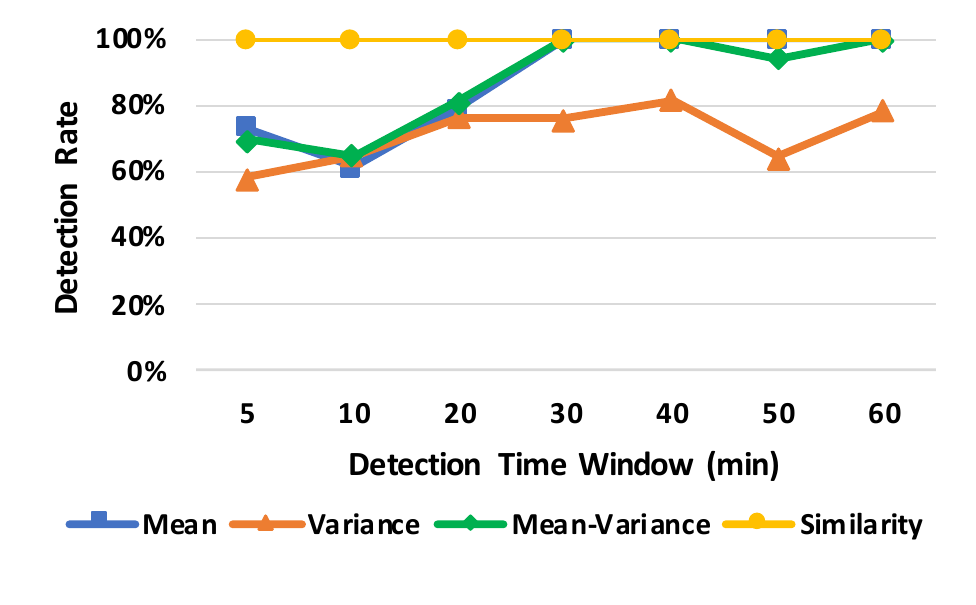}
    \vspace{-2mm}
    \caption{Detection Rates vs. Detection Time Window for SMO SVM}
    \label{fig:detectionratesvm}
    \vspace{-2mm}
\end{figure}

\subsection{Defense Effectiveness}

We use a popular ML tool Weka V3.8.6 to train the ML model and evaluate the effectiveness of the proposed defense scheme on the false data injection attack against the temperature sensor. Two classic ML algorithms, Weka J48 decision tree and Weka SMO support vector machines (SVM), are used.
We gathered the attack dataset $A$ and baseline dataset $B$ over a 24-hour period each from our real-world testbed.
We set the detection time window $t$ as $5$min, $10$min, $20$min, $30$min, $40$min, $50$min, and $60$min, respectively.
The detection rate is defined as the overall accuracy of recognizing feature vectors labeled as {\em attack} and {\em no attack} in the testing dataset.\looseness=-1

Figure \ref{fig:detectionratej48} shows the detection rate for each detection time window with the J48 decision tree based model. The proposed probability distribution similarity feature works better than mean, variance and (mean, variance) features when the detection time window is less than or equals to $20$ min. The defender who detects the attack would prefer a small detection time window so as to discover the attack quickly. An attack may not last long either.
In the case of the SVM-based ML model, the probability distribution similarity feature always outperforms other features, and its corresponding detection rate reaches 100\% with each detection time window as shown in Figure \ref{fig:detectionratesvm}.
It can be observed that our defense scheme is effective, and the proposed probability distribution similarity feature works better than the mean and variance related features.


%% file: sections/sec8_Conclusion.tex
\vspace{-2mm}
\section{Conclusion}
\label{conclusion}
\vspace{-2mm}

In this paper we demonstrate the feasibility and impact of a false data injection attack against KNX based building automation systems. In such an attack, an attacker may physically remove a KNX sensor, and launch a Man-In-the-Middle attack with the intent to inject false data into the BAS. 
We carefully analyze the impact of false data injection attacks on HVAC systems, and quantify the additional energy cost incurred by the false data injection attack by simulating two attack strategies.
We are the first to study the false data injection attack against a BAS and its impact. Machine learning-based strategies are presented to detect the false data injection attack based on features of the telegram inter-arrival time. The SVM classifier can achieve a detection rate of 100\% using the proposed probability distribution similarity feature as compared to the mean and variance related features with small detection time windows.

